# Dimensionality-induced change in topological order in multiferroic oxide superlattices

Megan E. Holtz[1,2], Elliot S. Padgett[1], Rachel Steinhardt[2], Charles M. Brooks[2], Dennis Meier[3], Darrell G. Schlom[2,4,5], David A. Muller[1,4], Julia A. Mundy[1,6]

[1] School of Applied and Engineering Physics, Cornell University, Ithaca, New York 14853, USA

[2] Department of Materials Science and Engineering, Cornell University, Ithaca, New York 14853, USA

[3] Department of Materials Science and Engineering, Norwegian University of Science and Technology, NTNU, 7491 Trondheim, Norway

[4] Kavli Institute at Cornell for Nanoscale Science, Ithaca, New York 14853, USA

[5] Leibniz-Institut für Kristallzüchtung, Max-Born-Str. 2, 12489 Berlin, Germany

[6] Department of Physics, Harvard University, Cambridge, Massachusetts 02138 USA

We construct ferroelectric $(LuFeO_3)_m/(LuFe_2O_4)$ superlattices with varying index *m* to study the effect of confinement on topological defects. We observe a thickness-dependent transition from neutral to charged domain walls and the emergence of fractional vortices. In thin $LuFeO_3$ layers, the volume fraction of domain walls grows, lowering the symmetry from *P*6$_3$*cm* to *P*3*c*1 before reaching the non-polar *P*6$_3$/*mmc* state, analogous to the high-temperature ferroelectric to paraelectric transition. Our study shows how dimensional confinement stabilizes textures beyond those in bulk ferroelectric systems.

Understanding transitions between a disordered and ordered phase was a major triumph of 20th century physics. In non-adiabatic transitions, the Kibble-Zurek framework describes a transition in which spontaneous symmetry breaking in disconnected regions creates topological defects [1,2]. While this model was developed in cosmology, it found application to a variety of solid-state systems, playing a central role in understanding phase transitions ranging from superfluid $^4$He and high-temperature superconductors. In particular, ferroelectric materials have topological defects such as vortices and domain walls, which have been used to study otherwise inaccessible topological phenomena in the same universality, answering cosmology-related questions [3–5]. The topological defects can be imaged at the micron scale with scanning probe or optical microscopy [6–9] or at the atomic-scale with transmission electron microscopy [10–12]. In addition to their significance for fundamental research, these vortices and domain walls exhibit emergent functional properties, representing nanoscale objects with distinct insulating, conducting, or magnetic properties that are not present in the homogeneous bulk phases [7,8,12–14].

Topologically rich structures in perovskite ferroelectric systems have recently been created and manipulated using geometric confinement to tune the interplay between strain and depolarization fields. Nanostructured systems such as ferroelectric disks, rods, and composites have displayed vortices, skyrmions, and waves [15–18]. Precise epitaxial growth can further generate new metastable phases hidden in the energy landscape and has been recently used to form ferroelectric/paraelectric superlattices that generate ferroelectric vortices [19] and polar skyrmions [20]. This opened the door to studying the chirality, negative capacitance, and piezo-electric responses in these topological structures [21–23]. These studies focused on "soft" ferroelectrics, where the spontaneous polarization rotates from the direction it has in the bulk structure in response to geometric confinement.

Here, we present atomically-precise $(LuFeO_3)_m/(LuFe_2O_4)_1$ superlattices synthesized by reactive-oxide molecular beam epitaxy [24] as a unique synthetic construction to manipulate the existing topological textures in uniaxial ferroelectrics. Hexagonal $LuFeO_3$ is an improper ferroelectric that is isostructural to a class of materials including hexagonal manganites, gallates, indates, and tungsten bronzes, and has a robust ferroelectric polarization for temperatures up to 1020 K [25–28]. These materials have domain walls and six-fold trimerization vortices in bulk systems. We layer this with $LuFe_2O_4$, which is epitaxially matched to $LuFeO_3$ but is non-polar in the bulk [29,30], and has a ferrimagnetic moment creating a room-temperature multiferroic as previously reported [24]. We use scanning transmission electron microscopy (STEM) to measure the polar displacements and the improper order parameter, [24,31] and map the domain walls and the underlying energy landscape. These measurements show that varying the thickness of $LuFeO_3$ changes the topological ordering and symmetries present in the system.

As shown in Fig. 1a, the improper ferroelectric polarization in $LuFeO_3$ is driven by a tilting of the iron-oxygen trigonal bipyramids which results in a polar "up-up-down" (+$P$) or "down-down-up" (-$P$) displacement of the lutetium atoms [32,33]. The combination of the trimerization (breaking $Z_3$ symmetry) and polar distortion (breaking $Z_2$ symmetry) leads to a net $Z_6$ symmetry. The primary order parameter responsible for the symmetry breaking can be described by the amplitude of the distortion, $Q$, and phase $\Phi$, which takes one of six discrete values $\Phi_n = \frac{n\pi}{3}$ ($n = 0, ..., 5$) in the polar $P6_3cm$ phase [26,34–36]. A lower symmetry state exists where the phase $\Phi$ varies continuously ($P3c1$), while the higher symmetry state with no distortions ($Q = 0$) gives the nonpolar $P6_3/mmc$ state [36]. The polarization arises due to a coupling to $Q$ ($P \sim Q^3 \cos 3\Phi$) [33,34,37].

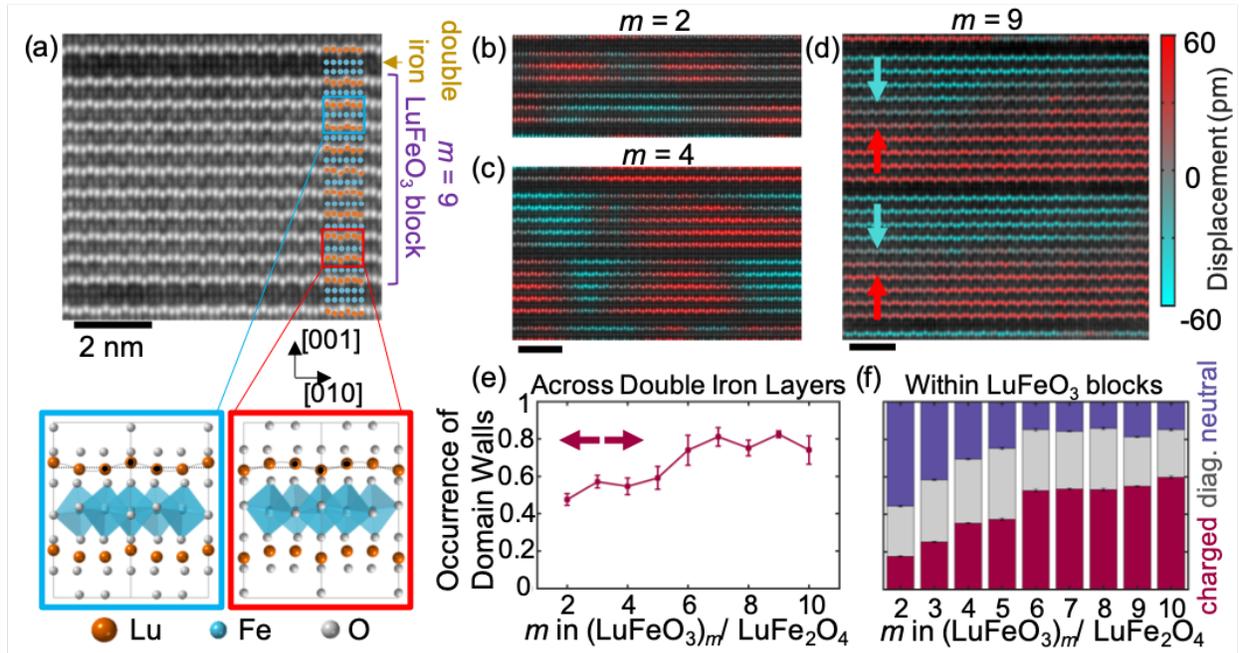

**Figure 1.** (a) HAADF-STEM image of an $m = 9$ (LuFeO$_3$)$_m$/LuFe$_2$O$_4$ superlattice, where bright (dark) contrast indicates lutetium (iron) atomic columns. Below is a cartoon of the crystal structure. A sinusoidal curve fits the lutetium displacements, giving the amplitude $Q$ and phase $\Phi$. (b-d): Polarization color overlay, with cyan indicating polarization down and red indicating polarization up, for $m = 2$ (b), 4 (c) and 9 (d), where head-to-head walls fall in the middle of the LuFeO$_3$ block. (e) Statistics showing a tail-to-tail configuration across the double iron layer is increasingly prevalent with larger $m$. (f) Occurrence of neutral walls (angles >60° from horizontal), charged walls (angles <30°), and diagonal walls (30° - 60°) within the LuFeO$_3$ block.

Figure 1b-d shows the polarization as color overlays as the thickness, $m$, of the LuFeO$_3$ is increased (SI Fig. 1 shows $P$, $Q$, and $\Phi$ images). The index $m$ corresponds to the number of formula-unit-thick LuFeO$_3$ layers in each repeat of the superlattice. Consistent with our previous report [24], we observe polar distortions for $m \geq 2$. The domains shown in Fig. 1c for low $m$ are small with a mixture of neutral and charged walls. As the thickness of the LuFeO$_3$ layer is increased, the domain structure becomes more coherent. In the $m = 9$ sample in Fig. 1d, there are consistent polarization down domains at the top of the LuFeO$_3$ block, and polarization up

domains at the bottom (for larger field of view images, see SI Fig. 2). This pattern at high *m* enforces a tail-to-tail polarization configuration (←→) across the LuFe$_2$O$_4$ layer, and a head-to-head domain wall (→←) confined within the LuFeO$_3$ block.

Charged polarization configurations are typically energetically costly and only appear in bulk hexagonal manganites because of the topologically protected six-fold vortices. No vortices are observed that would prevent the material from only forming energetically favorable neutral walls, yet the charged head-to-head walls persist at high *m*, indicating that six-fold vortices are not imperative to form charged domain walls in this structure. Based on EELS measurements (SI Fig. 3), we conclude that screening of the domain wall bound charges is achieved by varying the iron valence across this structure. Density functional theory has further predicted that the charged domain wall pattern would stabilize with increasing LuFeO$_3$ thickness due to hole transfer to the double iron layer [24]. The electrostatics in this confined system appear to generate the charged domain wall pattern.

We performed analysis of over 14,000 nm$^2$ (142,640 atomic columns) to generate robust statistics over *m* = 1 to 10. Consistent with the qualitative observations, the tail-to-tail polarization configuration is formed across most of the double iron layers (Fig. 1e), with increasing regularity for increasing *m*. The consequence is a propensity for head-to-head walls in the LuFeO$_3$ block, with higher regularity for higher *m* (SI Fig. 4). Further, charged head-to-head domain walls are stabilized in the thicker layers, while neutral domain walls are more prevalent in thinner layers (Fig. 1f). This regular domain architecture is not observed in thin LuFeO$_3$ grown between paraelectric layers of InFeO$_3$ (SI Fig. 5) or in ultra-thin epitaxial films [38], which exhibit suppression of ferroelectric order due to clamping at the interface. The (LuFeO$_3$)$_m$/(LuFe$_2$O$_4$) synthetic construct thus provides us with an experimental system that stabilizes and confines charged domain walls, which is distinct from simply considering ultrathin ferroelectric layers.

Analysis of the STEM images allows us to measure the primary order parameter ($Q$, $\Phi$) that drives the polarization configuration in the $(LuFeO_3)_m/(LuFe_2O_4)$ superlattices [12]. Figure 2a shows the six trimerization domains corresponding to the $Z_6$ symmetry. The HAADF-STEM images are overlaid with the color scheme in Fig. 2 and 3, for larger and smaller $m$, respectively. Similar to bulk hexagonal manganites, we observe a phase change $\Delta\Phi = \pm \pi/3$ at domain walls within the $LuFeO_3$ blocks. We do not observe a correlation of the phase across the $LuFe_2O_4$ beyond enforcing a tail-to-tail polarization configuration (SI Fig. 4), so each $LuFeO_3$ block can be considered as an isolated, quasi-two-dimensional system of determined thickness.

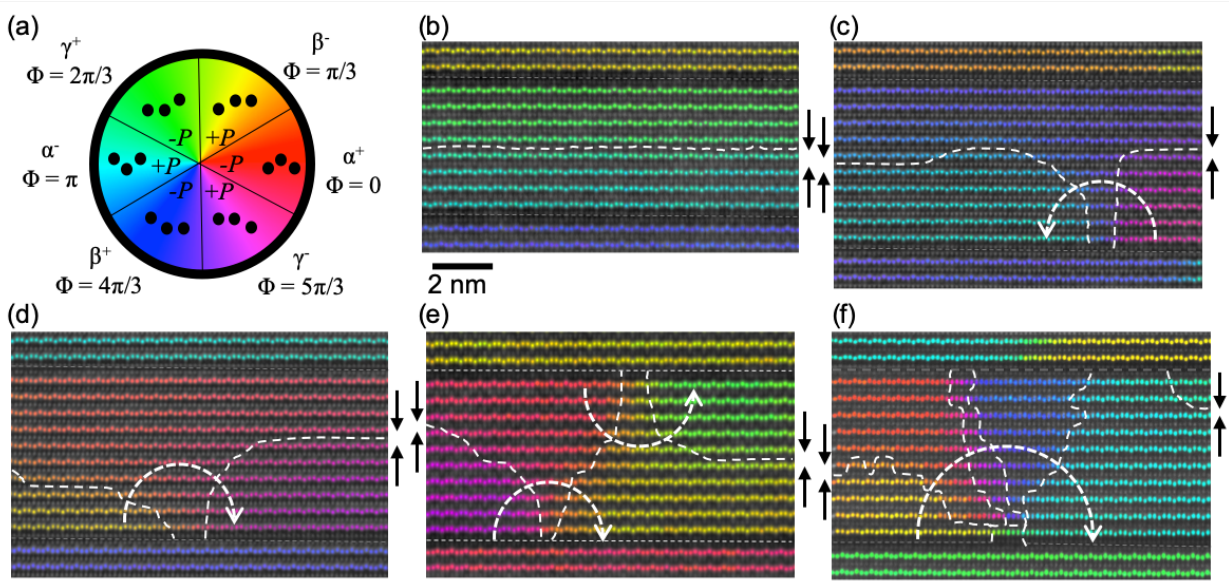

**Figure 2.** $\Phi$ overlay of STEM images showing the domain structure in $(LuFeO_3)_m/(LuFe_2O_4)$ superlattices for large $m$ ($m = 7,9$). (a) Cartoon of the projection of lutetium positions for different $\Phi$, with the color corresponding to the color overlays. (b) A stable charged domain wall, with black arrows showing polarization directions. (c-d) At the end of a domain in-plane, the phase rotates clockwise or anti-clockwise (white arrows), forming half vortices and anti-vortices to maintain the charged domain wall configuration. (e) For short in-plane domains, the phase can first wrap on one neighboring double iron layer and then the other. (f) Rarely, a vortex with five domains can be observed.

In our images of superlattices with $m > 4$, such as shown in Fig. 2, we observe "half-vortices" composed of three out of the six possible domain states sketched in Fig. 2a. The "core" of such fractional vortices is pinned to the double iron layers as displayed in Fig. 2c-d. In Fig. 2e, two half-vortices appear side-by-side, with phases wrapping in the same direction. The systematic formation of such half-vortices in $(LuFeO_3)_m/(LuFe_2O_4)$ allows the system to stabilize head-to-head walls within the $LuFeO_3$ blocks, while keeping a tail-to-tail configuration across the $LuFe_2O_4$ layers. This behavior is fundamentally different from analogous bulk systems where the structural trimerization enforces six-fold vortices. While the splitting of structural vortices into "fragmented vortices" has been predicted to occur in systems away from the ground state [39], the observed correlation between preferred domain wall orientations and half-vortices is unexpected. Rarely, five domains come together at a point, shown in Fig. 2f, possibly stabilized by defects. The electrostatics which drive the tail-to-tail polarization configuration across the $LuFe_2O_4$ layer discourage or prohibit the formation of a full vortex. Interestingly, as the phases wrap around the fractional vortices, they progress from $Z_6$ symmetry towards a U(1) symmetry near the core (SI Fig. 6), analogous to the six-fold vortices in hexagonal manganites [12].

In the bulk case, the primary order parameter related to the structural symmetry breaking drives the formation of vortices, which enforce electrostatically unfavorable charged ferroelectric domain walls. In this system, the charged domain walls appear to co-determine the topological feature formation, favoring fractional vortices. The appearance of these "fractional" vortices suggests that the impact of electrostatics in the $(LuFeO_3)_m/LuFe_2O_4$ superlattices is a stronger influence than in the isostructural bulk system, softening the rigid hierarchy of energy scales.

In thinner layers, as the confinement of the $LuFeO_3$ layer is increased, small domains with neutral domain walls are increasingly prevalent. Images of the $m = 2$ and 4 structures are shown in Fig. 3. For *m* = 4 in Fig. 3a, there are charged domain walls and fractional vortices, but also stripe patterns where the phase progresses across the in-plane direction of the sample. Tail-to-tail polarization orientation is not always enforced across the double iron layer. For *m* = 2 in Fig. 3b, we observe more instances of the topological stripe formation (which has a gradual phase winding as shown in SI Fig. 7), although charged domain walls can also be found. Both *m* = 2 and 4 show small domains in close proximity.

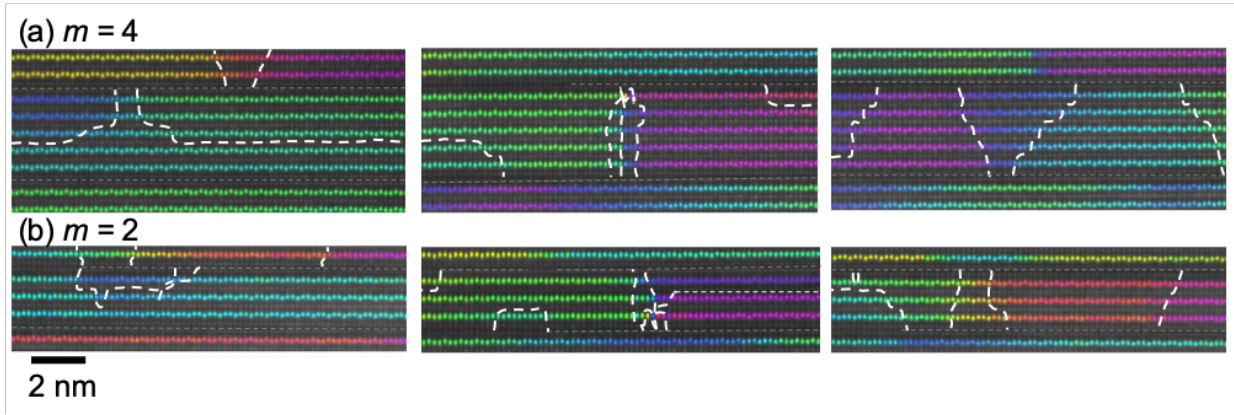

**Figure 3.** The order parameter $\Phi$ as a color overlay on the HAADF-STEM images for small *m* (*m* = 4, 2) where the neutral domain walls are observed with more equal occurrence. (a) for *m* = 4, we see a fairly equal co-existence of charged ferroelectric domain walls (left), and neutral + diagonal domain walls forming stripe patterns (right). We also see irregular numbers of domains coming together to a point in the middle image. (b) In the *m* = 2 case, we see smaller areas of charged domain walls (left) and more neutral walls forming in a stripe pattern (right). We also observe unusual numbers of domains intersecting (middle).

There is a resulting change to the global symmetry as the topological defects occupy an increasing fraction of the LuFeO$_3$ material with decreasing *m*. Figure 4a displays histograms of the logarithm of the occurrences of the structural order parameter, which maps the free energy landscape [12]. For *m*≥4, the domains have a well-defined energy landscape with six minima and $Z_6$ (*P6$_3$cm*) symmetry as expected. Here, intermediate states between the six wells corresponding to the domain walls. For *m* = 1, we observe the paraelectric state, corresponding to *P6$_3$/mmc* symmetry. Interestingly, for *m* = 2, we observe a fairly uniform distribution of the structural order parameter, which we might expect for a state with $P3c1$ symmetry, with values at low $Q$ indicating some contribution from paraelectric states. Likewise, the $m = 3$ state shows slightly more weight in the ferroelectric wells, while also showing some contribution from paraelectric states. These distributions are integrated from many images and reflect the non-uniformity of the sample.

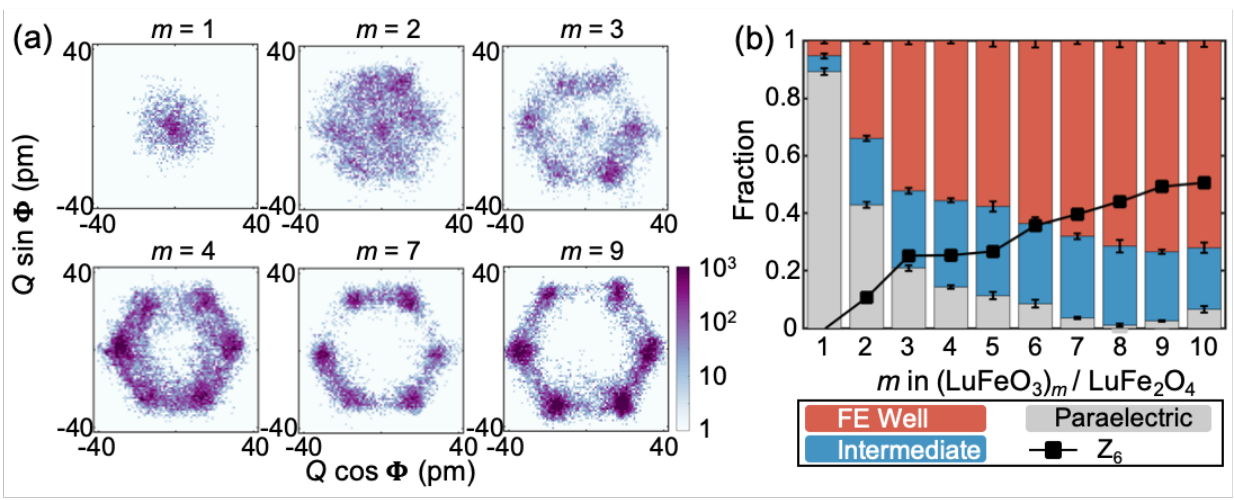

**Figure 4.** (a) Histograms of the order parameter for different $m$ in $(LuFeO_3)_m/(LuFe_2O_4)$ superlattices, with the logarithm of the occurrences plotted. $m = 1$ shows a parabolic well, consistent with a paraelectric order parameter distribution. For $m = 2$ and 3, $Q$ is slightly larger and the values of $\Phi$ are widely spread, indicating a combination of ferroelectric domain and intermediate $\Phi$ states. For $m = 4$, the six ferroelectric domains are visible while maintaining a large fraction of intermediate states. For $m = 7$ and 9, the six ferroelectric domains comprise the majority of the observed states, and the $Q$ value is higher. (b) Relative amount of states in ferroelectric (FE) wells vs. in the intermediate states (between wells) vs. the paraelectric state. The difference between the states in the well and in the intermediate states tracks the relative amount of $Z_6$ symmetry.

As domain walls overlap within 2-3 nm of a vortex core in hexagonal $ErMnO_3$, $U(1)$ symmetry emerges [12]. In these superlattices, the confinement length for $m = 2$ and 3 is similar to the domain wall width (SI Fig. 8), and the domain walls similarly overlap. For the $m = 2$ case, we observe local, atomic-scale polar distortions, but due to the abundance of domain walls, an overall $P3c1$ symmetry. This is reminiscent of bulk ferroelectric transitions observed in $YMnO_3$, where above the ferroelectric transition temperature, the average structure is $P6_3/mmc$ symmetry, while pair distribution analysis suggests that there are local fluctuations that lower the local symmetry to $P3c1$ [36]. The transition with confinement is quantified in Fig. 4b. We observe a cross-over in the dominant occupied states: the paraelectric state is dominant for $m = 1$; the paraelectric, ferroelectric well, and in-between well "intermediate" states are roughly balanced for $m = 2$; and the ferroelectric well states are most prevalent for $m > 3$.

In conclusion, we have demonstrated atomic-scale control of domain wall placement in the $(LuFeO_3)_m/(LuFe_2O_4)$ superlattice system, which further provides insight into the topology and symmetry of uniaxial ferroelectrics under dimensional confinement. In the thicker $LuFeO_3$ blocks, the topological defects consist largely of charged domain walls with "fractional" vortices pinned on the boundary $LuFe_2O_4$ layers—suggesting that the electrostatics imposed by the superlattice drive charged domain wall formation and the resulting three-fold vortices. This charged domain wall pattern is disrupted for thinner $LuFeO_3$ blocks, where smaller domains with neutral domain walls prevail. As the domain walls comprise more of the material, an emergent $P3c1$ symmetry is observed for $m = 2$ before becoming paraelectric for $m = 1$. We image this transition with atomic-scale resolution, observing both the local displacements and the average symmetry from sampling the overall energy landscape. This provides direct and simultaneous imaging of phase transition in confined $LuFeO_3$ layers. Moreover, as the vortex cores in these materials were previously associated with fractional electronic charge and quantized magnetic flux, highly confined vortex and fractional vortex states could display novel functionality.

**Acknowledgements**

We acknowledge discussions with Hena Das, Elizabeth Nowadnick, Craig Fennie, Sverre Selbach, Andres Cano, Sang-Wook Cheong, and Chris Nelson. We acknowledge technical support with the electron microscopy from Earl Kirkland, Malcolm Thomas, John Grazul and Mariena Silvestry Ramos. Research was supported by the US Department of Energy, Office of Basic Energy Sciences, Division of Materials Sciences and Engineering, under Award No. DE-SC0002334. The electron microscopy studies made use of the electron microscopy facility of the Cornell Center for Materials Research, a National Science Foundation (NSF) Materials Research Science and Engineering Centers program (DMR-1719875). DM was supported by NTNU via the Onsager Fellowship Program and the Outstanding Academic Fellows Program.

**References**

[1] T. W. B. Kibble, J. Phys. A. Math. Gen. **9**, 1387 (1976).
[2] W. H. Zurek, Nature **317**, 505 (1985).
[3] S. M. Griffin, M. Lilienblum, K. T. Delaney, Y. Kumagai, M. Fiebig, and N. A. Spaldin, Phys. Rev. X **2**, 1 (2012).
[4] S.-Z. Lin, X. Wang, Y. Kamiya, G.-W. Chern, F. Fan, D. Fan, B. Casas, Y. Liu, V. Kiryukhin, W. H. Zurek, C. D. Batista, and S.-W. Cheong, Nat. Phys. **10**, 970 (2014).
[5] Q. N. Meier, M. Lilienblum, S. M. Griffin, K. Conder, E. Pomjakushina, Z. Yan, E. Bourret, D. Meier, F. Lichtenberg, E. K. H. Salje, N. A. Spaldin, M. Fiebig, and A. Cano, Phys. Rev. X **7**, 041014 (2017).
[6] M. Fiebig, T. Lottermoser, D. Fröhlich, A. V. Goltsev, and R. V. Pisarev, Nature **419**, 818 (2002).
[7] T. Choi, Y. Horibe, H. T. Yi, Y. J. Choi, W. Wu, and S.-W. Cheong, Nat. Mater. **9**, 253 (2010).
[8] D. Meier, J. Seidel, a. Cano, K. Delaney, Y. Kumagai, M. Mostovoy, N. a. Spaldin, R. Ramesh, and M. Fiebig, Nat. Mater. **11**, 284 (2012).
[9] W. Wu, Y. Horibe, N. Lee, S.-W. Cheong, and J. R. Guest, Phys. Rev. Lett. **108**, 077203 (2012).
[10] Q. H. Zhang, L. J. Wang, X. K. Wei, R. C. Yu, L. Gu, A. Hirata, M. W. Chen, C. Q. Jin, Y. Yao, Y. G. Wang, and X. F. Duan, Phys. Rev. B - Condens. Matter Mater. Phys. **85**, 2 (2012).
[11] T. Matsumoto, R. Ishikawa, T. Tohei, H. Kimura, Q. Yao, H. Zhao, X. Wang, D. Chen, Z. Cheng, N. Shibata, and Y. Ikuhara, Nano Lett. **13**, 4594 (2013).
[12] M. E. Holtz, K. Shapovalov, J. A. Mundy, C. S. Chang, Z. Yan, E. Bourret, D. A. Muller, D. Meier, and A. Cano, Nano Lett. **17**, 5883 (2017).
[13] Y. Geng, N. Lee, Y. J. Choi, S. W. Cheong, and W. Wu, Nano Lett. **12**, 6055 (2012).
[14] J. A. Mundy, J. Schaab, Y. Kumagai, A. Cano, M. Stengel, I. P. Krug, D. M. Gottlob, H. Doğanay, M. E. Holtz, R. Held, Z. Yan, E. Bourret, C. M. Schneider, D. G. Schlom, D. A. A. Muller, R. Ramesh, N. A. A. Spaldin, and D. Meier, Nat. Mater. **16**, 622 (2017).
[15] Z. Liu, B. Yang, W. Cao, E. Fohtung, and T. Lookman, Phys. Rev. Appl. **8**, 034014 (2017).
[16] Y. Nahas, S. Prokhorenko, L. Louis, Z. Gui, I. Kornev, and L. Bellaiche, Nat. Commun. **6**, 1 (2015).

[17] I. I. Naumov, L. Bellaiche, and H. Fu, Nature **432**, 737 (2004).
[18] A. Schilling, D. Byrne, G. Catalan, K. G. Webber, Y. A. Genenko, G. S. Wu, J. F. Scott, and J. M. Gregg, Nano Lett. **9**, 3359 (2009).
[19] A. K. Yadav, C. T. Nelson, S. L. Hsu, Z. Hong, J. D. Clarkson, C. M. Schlepütz, A. R. Damodaran, P. Shafer, E. Arenholz, L. R. Dedon, D. Chen, A. Vishwanath, A. M. Minor, L. Q. Chen, J. F. Scott, L. W. Martin, R. Ramesh, C. M. Schlepüetz, A. R. Damodaran, P. Shafer, E. Arenholz, L. R. Dedon, D. Chen, A. Vishwanath, A. M. Minor, L. Q. Chen, J. F. Scott, L. W. Martin, and R. Ramesh, Nature **530**, 198 (2016).
[20] S. Das, Y. L. Tang, Z. Hong, M. A. P. Gonçalves, M. R. McCarter, C. Klewe, K. X. Nguyen, F. Gómez-Ortiz, P. Shafer, E. Arenholz, V. A. Stoica, S. L. Hsu, B. Wang, C. Ophus, J. F. Liu, C. T. Nelson, S. Saremi, B. Prasad, A. B. Mei, D. G. Schlom, J. Íñiguez, P. García-Fernández, D. A. Muller, L. Q. Chen, J. Junquera, L. W. Martin, and R. Ramesh, Nature **568**, 368 (2019).
[21] A. R. Damodaran, J. D. Clarkson, Z. Hong, H. Liu, A. K. Yadav, C. T. Nelson, S.-L. Hsu, M. R. McCarter, K.-D. Park, V. Kravtsov, A. Farhan, Y. Dong, Z. Cai, H. Zhou, P. Aguado-Puente, P. García-Fernández, J. Íñiguez, J. Junquera, A. Scholl, M. B. Raschke, L.-Q. Chen, D. D. Fong, R. Ramesh, and L. W. Martin, Nat. Mater. **16**, 1003 (2017).
[22] P. Shafer, P. García-Fernández, P. Aguado-Puente, A. R. Damodaran, A. K. Yadav, C. T. Nelson, S.-L. Hsu, J. C. Wojdeł, J. Íñiguez, L. W. Martin, E. Arenholz, J. Junquera, and R. Ramesh, Proc. Natl. Acad. Sci. **115**, 915 (2018).
[23] A. K. Yadav, K. X. Nguyen, Z. Hong, P. García-Fernández, P. Aguado-Puente, C. T. Nelson, S. Das, B. Prasad, D. Kwon, S. Cheema, A. I. Khan, C. Hu, J. Íñiguez, J. Junquera, L. Q. Chen, D. A. Muller, R. Ramesh, and S. Salahuddin, Nature **565**, 468 (2019).
[24] J. A. Mundy, C. M. Brooks, M. E. Holtz, J. A. Moyer, H. Das, A. F. Rébola, J. T. Heron, J. D. Clarkson, S. M. Disseler, Z. Liu, A. Farhan, R. Held, R. Hovden, E. Padgett, Q. Mao, H. Paik, R. Misra, L. F. Kourkoutis, E. Arenholz, A. Scholl, J. A. Borchers, W. D. Ratcliff, R. Ramesh, C. J. Fennie, P. Schiffer, D. A. Muller, and D. G. Schlom, Nature **537**, 523 (2016).
[25] E. Magome, C. Moriyoshi, Y. Kuroiwa, A. Masuno, and H. Inoue, Jpn. J. Appl. Phys. **49**, 09ME06 (2010).
[26] H. Das, A. L. Wysocki, Y. Geng, W. Wu, and C. J. Fennie, Nat. Commun. **5**, 2998 (2014).
[27] W. Wang, J. Zhao, W. Wang, Z. Gai, N. Balke, M. Chi, H. N. Lee, W. Tian, L. Zhu, X. Cheng, D. J. Keavney, J. Yi, T. Z. Ward, P. C. Snijders, H. M. Christen, W. Wu, J. Shen, and X. Xu, Phys. Rev. Lett. **110**, 1 (2013).
[28] S. M. Disseler, J. A. Borchers, C. M. Brooks, J. A. Mundy, J. A. Moyer, D. A. Hillsberry, E. L. Thies, D. A. Tenne, J. Heron, M. E. Holtz, J. D. Clarkson, G. M. Stiehl, P. Schiffer, D. A. Muller, D. G. Schlom, and W. D. Ratcliff, Phys. Rev. Lett. **114**, 217602 (2015).
[29] D. Niermann, F. Waschkowski, J. de Groot, M. Angst, and J. Hemberger, Phys. Rev. Lett. **109**, 16405 (2012).
[30] S. Lafuerza, J. García, G. Subías, J. Blasco, K. Conder, and E. Pomjakushina, Phys. Rev. B **88**, 85130 (2013).
[31] L. Li, P. Gao, C. T. Nelson, J. R. Jokisaari, Y. Zhang, S. J. Kim, A. Melville, C. Adamo, D. G. Schlom, and X. Pan, Nano Lett. **13**, 5218 (2013).
[32] B. B. Van Aken, T. T. M. Palstra, A. Filippetti, and N. A. Spaldin, Nat. Mater. **3**, 164 (2004).

[33] C. J. Fennie and K. M. Rabe, Phys. Rev. B - Condens. Matter Mater. Phys. **72**, 1 (2005).
[34] S. Artyukhin, K. T. Delaney, N. A. Spaldin, and M. Mostovoy, Nat. Mater. **13**, 42 (2014).
[35] M. Lilienblum, T. Lottermoser, S. Manz, S. M. Selbach, A. Cano, and M. Fiebig, Nat. Phys. **11**, 1070 (2015).
[36] S. H. Skjærvø, Q. N. Meier, M. Feygenson, N. A. Spaldin, S. J. L. Billinge, E. S. Bozin, and S. M. Selbach, Phys. Rev. X **9**, 31001 (2019).
[37] A. Cano, Phys. Rev. B - Condens. Matter Mater. Phys. **89**, 214107 (2014).
[38] J. Nordlander, M. Campanini, M. D. Rossell, R. Erni, Q. N. Meier, A. Cano, N. A. Spaldin, M. Fiebig, and M. Trassin, Nat. Commun. **10**, 5591 (2019).

**Materials and Methods**

Thin films of $(LuFeO_3)_m/(LuFe_2O_4)$ were grown by reactive-oxide molecular-beam epitaxy in a Veeco GEN10 system on (111) $(ZrO_2)_{0.905}(Y_2O_3)_{0.095}$ (or 9.5 mol% yttria-stabilized zirconia) substrates, as described in Ref [24].

Cross-sectional TEM specimens were prepared using an FEI Strata 400 Focused Ion Beam (FIB) with a final milling step of 2 keV to reduce surface damage. High-resolution HAADF-STEM images were acquired on a 100-keV Nion UltraSTEM, a fifth-order aberration-corrected microscope, and a 300-keV FEI Titan Themis.

The lutetium distortions were quantified from HAADF-STEM images. Several images were acquired with fast dwell time (<1-6 us) and averaged to reduce scan noise and the lutetium atomic positions were determined through segmentation using a threshold plus watershed algorithm, followed by two-dimensional Gaussian fitting procedure that was implemented in MATLAB. The position of each atom was then compared to its neighboring atom on each side in the same atomic plane, and the order parameters fit using the equation in Fig. 1a. To garner statistics for each layering, the order parameters, atomic positions, and layer type *m* were collected for over 14,000 nm$^2$ (data from 142,640 atomic columns) of material.